\documentclass[11pt,twoside]{article}
\usepackage{asp2010}
\usepackage{graphicx}
\usepackage{subfigure} 

\resetcounters

\begin{document}

\title{A GPU-based survey for millisecond radio transients using ARTEMIS}
\author{W.~Armour$^1$, A.~Karastergiou$^2$, M.~Giles$^1$, C.~Williams$^3$, A.~Magro$^4$, K.~Zagkouris$^2$, S.~Roberts$^5$, S.~Salvini$^3$, F.~Dulwich$^3$ and B.~Mort$^3$
\affil{$^1$Institute for the Future of Computing, Oxford Martin School, Oxford e-Research Centre, University of Oxford, Keble Road, OX1 3QG}
\affil{$^2$Astrophysics, University of Oxford, Denys Wilkinson Building, Keble Road,OX1 3RH}
\affil{$^3$Oxford e-Research Centre, University of Oxford, Keble Road, OX1 3QG}
\affil{$^4$Department of Physics, University of Malta, Msida MSD 2080, Malta}
\affil{$^5$Department of Engineering Science, University of Oxford, OX1 3PJ}}

\begin{abstract}
Astrophysical radio transients are excellent probes of extreme
physical processes originating from compact sources within our Galaxy and
beyond. Radio frequency signals emitted from these objects provide a
means to study the intervening medium through which they travel.
Next generation radio telescopes are designed to explore the vast
unexplored parameter space of high time resolution astronomy, but
require High Performance Computing (HPC) solutions to process the
enormous volumes of data that are produced by these telescopes. We have
developed a combined software/hardware solution (code named ARTEMIS) for
real-time searches for millisecond radio transients, which uses GPU technology
to remove interstellar dispersion and detect millisecond radio bursts from
astronomical sources in real-time. Here we present an introduction to ARTEMIS.
We give a brief overview of the software pipeline, then focus specifically on
the intricacies of performing incoherent de-dispersion.
We present results from two brute-force algorithms. The first is a GPU
based algorithm, designed to exploit the L1 cache of the NVIDIA Fermi GPU. Our
second algorithm is CPU based and exploits the new AVX units in Intel Sandy
Bridge CPUs.
\end{abstract}

\section{Introduction}
ARTEMIS stands for {\it Advanced Radio Transient Event Monitor and
Identification System}. It is a project being carried out at Oxford
(Astrophysics, OeRC, Engineering Science) aimed at the real-time processing of
high-time resolution data from radio astronomy (Karastergiou et al. in
preparation). 
The project aims to develop the software and piece together the hardware for
surveys of fast transients and pulsars using next generation radio telescopes
such as LOFAR and MeerKAT.
Real-time processing is essential to ensure that broadband data streams are
reduced to manageable rates both for storage and further processing. The
ARTEMIS servers perform (in real-time) all the operations necessary to
discover short duration radio pulses from pulsars and fast transients
\citep{wbd+11}, thanks to a modular software structure operating in a C++
scalable framework (PELICAN, developed at the OeRC).
AMPP (ARTEMIS Modular PELICAN Pipelines) is the software that we have
developed for receiving the data, for further channelisation in finer
frequency, generation of Stokes parameters, excision of radio frequency
interference, integration, real-time dispersion searches and detection of
interesting signals across multiple telescopes, in high-throughput CPU and GPU
code. This article describes the results of the GPU implementation of the
real-time incoherent de-dispersion aspect and gives a comparison to a CPU
implementation. Recently Magro et. al. \citep{alessio} published a GPU
implementation of incoherent de-dispersion (MDSM). Here we present a new,
optimised kernel, which we have used with the MDSM wrapper.

\section{Searching for radio transients}
\begin{figure}[b!]
\begin{center}
\includegraphics[scale = 0.30]{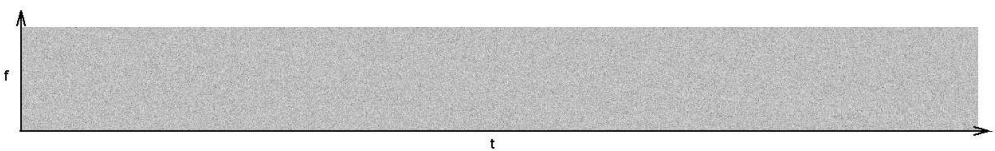}\label{fg:fakedata}
%\vspace{-0.2in}
\caption{An example of a test data-set (total intensity as a function of
$(f,t)$) for 496 channels. A weak signal can be seen moving from top
left to bottom right.}
%\vspace{-0.1in}
\end{center}
\end{figure}

The quadratic cold plasma dispersion law of the interstellar medium
results in radio pulses at lower frequencies arriving at Earth later
than their high frequency counterparts \citep{lk05}. To take advantage
of the fact that astrophysical radio bursts are typically broadband,
integration over frequency is essential to increase signal to
noise. Incoherent de-dispersion is the process of shifting the (power) data
inside each individual frequency channel to counter the effect of interstellar
dispersion before frequency integration. Given the quadratic relationship
between time delay and frequency, the phenomenon is governed by a single free
parameter, known as the dispersion measure (DM), which is the integrated
electron number density along the line of sight to the source. Figure
\ref{fg:fakedata} shows simulated filterbank data (using the sigproc package),
with a dispersed radio signal sitting in noise.
In a blind search for dispersed radio bursts, the DM is unknown. A
large range of DM values is typically searched, by shifting each
frequency channel by the appropriate amount of time for each DM being
searched. This results in each data point (in the frequency-time
domain) being used many times for all the dispersion curves that it
contributes to, a useful quality for GPU acceleration.

\section{Acceleration via GPU computing}
In order to produce a GPU kernel that can achieve a significant proportion of
the peak performance of the GPU we need to ensure three things. The first is
that the accumulator that stores the integrated value of the intensity (along
the trial dispersion curve) sits in the fastest area of memory. The second is
that the correct data from the $(f,t)$ domain is always available to
the streaming multiprocessors. The third is that the shifting value is
calculated using as few operations as possible.
\begin{figure}[t!]
\begin{center}
\mbox{
\subfigure{\includegraphics[scale=0.36]{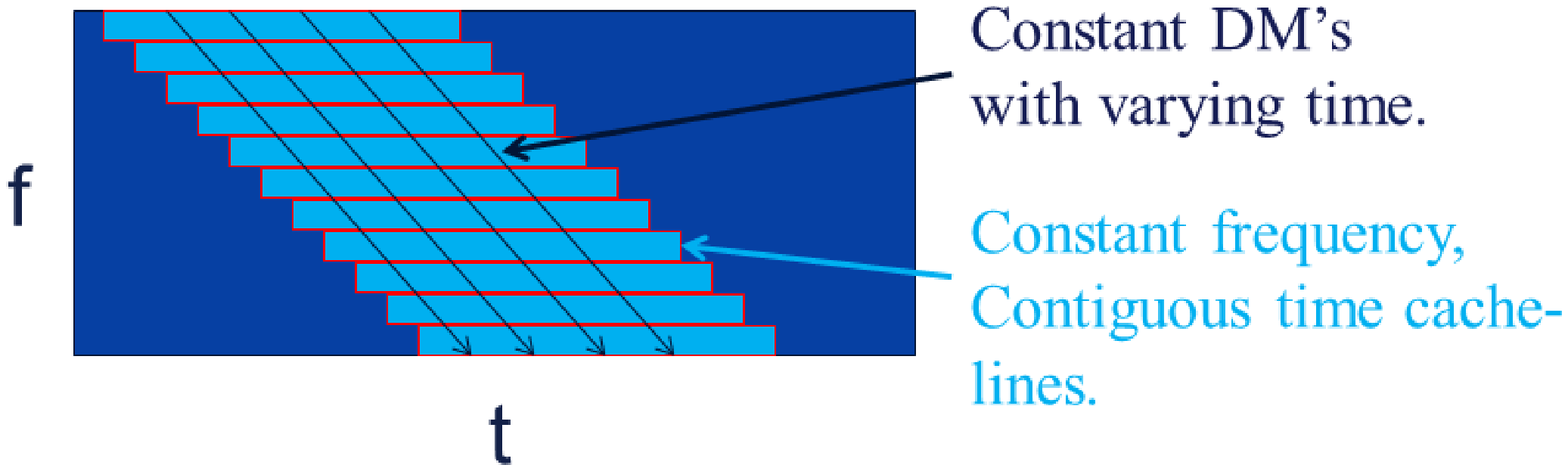}}\quad
\subfigure{\includegraphics[scale=0.4]{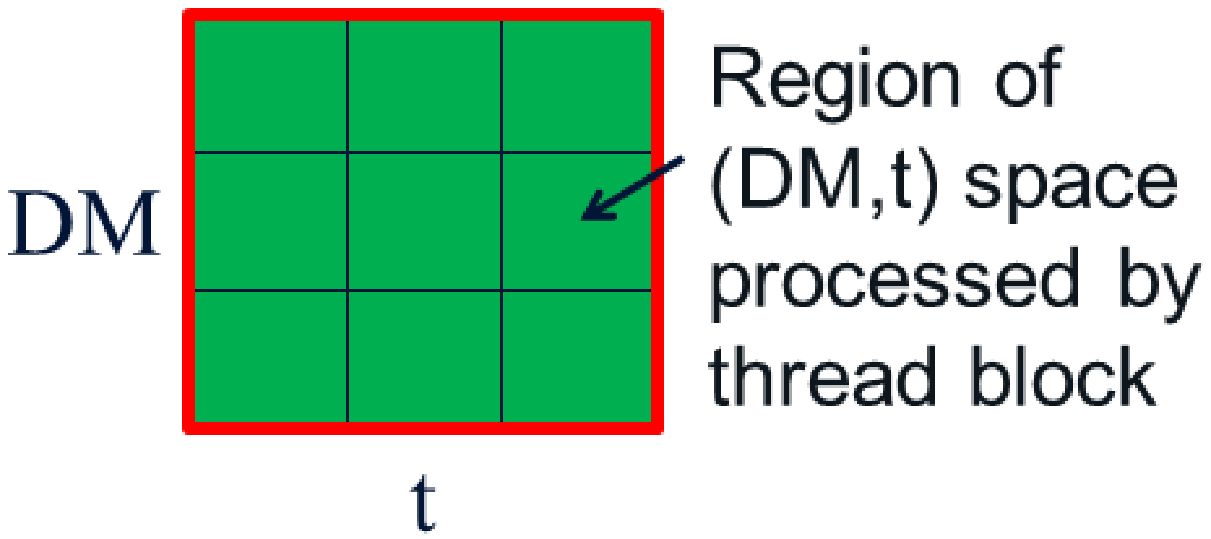}}
}
\caption{Left: A single thread loads data from cache-lines of constant
frequency, contiguous time and increments accumulators for multiple points in
$(DM,t)$ space. Right: Each thread-block processes an area of $(DM,t)$
space.}\label{fg:algorithm}
%\vspace{-0.2in}
\end{center}
\end{figure}
The GPU algorithm presented is designed to exploit the new fast L1 cache
present on the NVIDIA Fermi hardware. The algorithm is designed to reuse
cache-lines that are present in the L1 cache, vastly reducing the need to
transfer the same data from main graphics memory multiple times. This is
achieved by each thread processing several time elements for its given value
of dispersion, holding these values in local registers (Fig.
\ref{fg:algorithm}, left). This gives rise to each thread-block processing a
rectangular area of the dispersion-time, $(DM,t)$, space ensuring cache-lines
of $(f,t)$ data are reused multiple times (Fig. \ref{fg:algorithm}, right).

\section{Comparisons of GPU and CPU algorithms}
\begin{figure}[t!]
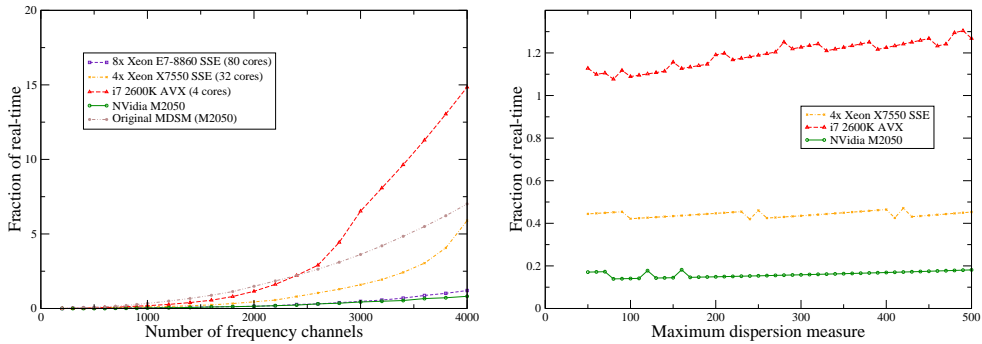

\begin{center}
\mbox{
\subfigure{\includegraphics[scale=0.27]{O02_f4l.eps}}\quad
\subfigure{\includegraphics[scale=0.27]{O02_f4r.eps}}
}
\caption{Left: Plot varying the number of channels showing execution time as
percentage of real-time. Here we hold the total number of trial dispersion
searches equal to the number of channels.
Right: Plot varying the maximum dispersion measure, whilst holding the total
number of trial dispersions constant (number of channels = number of trial
dispersion searches $=2000$).}\label{fg:hardware}
%\vspace{-0.2in}
\end{center}
\end{figure}

In this section we present results from our GPU kernel and compare these
results to a vector-parallel CPU code that exploits the SSE registers on a
multiprocessor Intel Xeon machine or the AVX registers on a new Intel i7
Sandy Bridge based machine (Overclocked from 3.2GHz to 4.2GHz, employing
1600MHz DDR3 SD-RAM). 
\begin{figure}[b!]
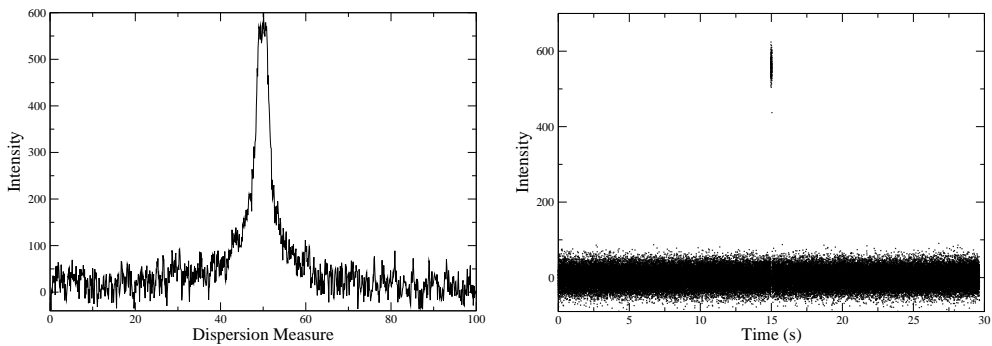

\begin{center}
\mbox{
\subfigure{\includegraphics[scale=0.27]{O02_f3l.eps}}\quad
\subfigure{\includegraphics[scale=0.27]{O02_f3r.eps}}
}
\caption{Left: The result of a dispersion search on simulated data with a DM
of $50$ pc cm$^{-3}$. Right: The square pulse signal recovered at the identified
DM has all the characteristics of the simulated signal.}\label{fg:detection}
%\vspace{-0.2in}
\end{center}
\end{figure}
The CPU code has been designed with maximum cache-line usage in mind and use
the Intel intrinsics in the vector parts of the code. Results from a
vectorised code using the Intel auto-vectorizer have not been presented
because they are consistently slower (approximately 3x, in our region of
interest) compared with our vectorised code.
Figure \ref{fg:detection} shows results from the CPU code, demonstrating
results that are in exact agreement with our simulated data.
In figure \ref{fg:hardware} (left) we present the proportion of real-time taken by
the CPU/GPU codes (including different platforms) against a varying number of
frequency channels. Importantly we hold the maximum dispersion measure at
$200$. However to ensure that we do not sub-sample the data we set the total
number of dispersion measures equal to the total number of channels.
Figure \ref{fg:hardware} (right) shows how the different codes perform as we
increase the maximum value of the dispersion measure with a fixed number of
frequency channels. Here we hold the number of dispersion measures equal to the
number of channels but we change the value of the equally spaced dispersion
curves. In both cases we observe better performance from the GPU code.
\section{Conclusions and future work}
With typical parameters of a dispersion search (~2000 frequency
channels and dispersion measures to search), we estimate our kernel
achieves approximately 40\% - 50\% of peak GPU performance. This leaves
little margins for improvement for GPU based, brute-force, incoherent
dedispersion algorithms and makes real-time dispersion searches a
possibility under many different observing situations. Our kernel has
been tested successfully in a real environment within ARTEMIS.
To try and achieve a balance between the CPU and GPU computing powers our
future work will focus on implementing vectorisation in the poly-phase filter
using AVX registers and the Intel intrinsics.

\bibliography{O02}

\end{document}